\documentclass[prc,reprint,onecolumn,amsmath,amssymb,superscriptaddress]{revtex4-1}
\usepackage{graphicx}
\newcommand{\be}{\begin{equation}}
\newcommand{\ee}{\end{equation}}
\newcommand{\beq}{\begin{eqnarray}}
\newcommand{\eeq}{\end{eqnarray}}
\newcommand{\ba}{\begin{array}}
\newcommand{\ea}{\end{array}}

\usepackage{color}

\begin{document}

\title{J/$\psi p$ Scattering Length from GlueX Threshold Measurements}

\date{\today}

%--------------------- AUTHOR LIST ---------------------------
\author{\mbox{Igor~Strakovsky}}
\altaffiliation{Corresponding author; \texttt{igor@gwu.edu}}
\affiliation{Institute for Nuclear Studies, Department of Physics, The George 
	Washington University, Washington, DC 20052, USA}

\author{\mbox{Denis~Epifanov}}
\affiliation{Budker Institute of Nuclear Physics SB RAS, Novosibirsk 
	630090, Russia}
\affiliation{Novosibirsk State University, Novosibirsk 630090, Russia}

\author{\mbox{Lubomir~Pentchev}}
\affiliation{Thomas Jefferson National Accelerator Facility, Newport 
	News, Virginia 23606, USA}

\noaffiliation

%---------------------- ABSTRACT -----------------------------
\begin{abstract}
The quality of the recent GlueX $J/\psi $ photoproduction data from Hall~D at Jefferson Laboratory 
and the proximity of the data to the energy threshold, gives access to a variety of interesting 
physics aspects. As an example, an estimation of the $J/\psi$-nucleon scattering length 
$\alpha_{J/\psi p}$ is provided  within the vector meson dominance model. It results in 
$|\alpha_{J/\psi p}| = (3.08\pm 0.55 ({\rm stat.}) \pm 0.42 ({\rm syst.}))$~mfm which is much 
smaller than a typical size of a hadron.
\end{abstract}

\maketitle

%----------------------- Introduction ------------------------
%\section{Introduction}
%\label{sec:intro}

Since the discovery of the $J/\psi(1S)$ resonance~\cite{Aubert:1974js,Augustin:1974xw}, it became an
attractive probe to study interactions with hadronic matter. The $J/\psi$ is a vector meson and our 
understanding of the coupling of the other vector mesons, such as $\omega$, $\rho$, and $\phi$ to
hadrons, has been strongly influenced by the photoproduction cross sections of these particles. In 
particular, the cross sections at small momentum transfer are characterized by an energy-independent 
exponential function of the square of the momentum transfer. 
%This simple empirical property has been interpreted as evidence for a diffractive-like production 
%mechanism. 
Explicit models, such as vector meson dominance (VMD), enable one to calculate the meson-hadron 
couplings from these cross sections. As a consequence, it is natural to study the photoproduction 
of the $J/\psi(1S)$ in order to compare it to the other vector mesons. The exclusive near-threshold 
photoproduction of charmonium allows for the study of the $J/\psi N$ interaction dominated by hard 
gluon exchange (due to the heavy charm quarks), thus providing a unique probe of the gluonic 
field in the nucleon at high $x$. The behavior of the cross section near threshold is related to 
the $J/\psi N$ scattering length, which can be used to study the binding of charmonium with the 
nucleon and nuclei. A large scattering length corresponds to a high binding energy. A small 
positive or negative scattering length $\alpha_{J/\psi N}$ may indicate a weakly repulsive or
attractive $J/\psi N$ interaction if there is no $J/\psi N$ bound state.

Recently, the GlueX Collaboration reported the first total cross section $\sigma_t$ measurements 
for the exclusive reaction $\gamma p\to J/\psi p$ at threshold~\cite{Ali:2019lzf}. This is a 
unique experiment that measured $\sigma_t(E)$ from threshold at a photon energy of $E_{\gamma} = 
8.2$~GeV to $11.85$~GeV. Previous old measurements at Cornell~\cite{Gittelman:1975ix} at $11$~GeV 
and SLAC~\cite{Camerini:1975cy} above $13$~GeV were inclusive and also on nuclear targets. Thanks 
to the full acceptance of the GlueX detector for this reaction, this measurement avoids 
uncertainties in the determination of $\sigma_t$ from the forward differential cross sections 
that affected the SLAC experiment. The GlueX experiment uses tagged real photons produced from 
12~GeV electrons by coherent Bremsstrahlung on a thin diamond radiator. The coherent peak was set 
at $E_\gamma$ right above the threshold, $8.2-9$~GeV, allowing to do measurements very close to the 
threshold where the cross-section vanishes.  The full acceptance of the GlueX detector is achieved 
by means of a solenoidal magnet with central and forward tracking systems and a barrel 
electromagnetic calorimeter, all inside the magnet, and additional calorimeter and time-of-flight 
systems covering the detector in the forward direction.

%In this work, we will report our determination of the $J/\psi p$ scattering length using recent 
%charmonium photoproduction data on the proton that was conducted by the GlueX Collaboration at 
%Jefferson Lab.

%------------Scattering Length Determination -----------------------
%\section{Scattering Length Determination}
%\label{sec:SL}

Near-threshold cross sections of good accuracy allow the extraction of various useful parameters, 
in particular, resonance masses and scattering lengths, see, for instance, Ref.~\cite{Strakovsky:2014wja}.  
In general, the total cross section for an inelastic reaction $ab\to cd$ with particle masses $m_a + m_b 
< m_c + m_d$ can be written as $\sigma_t = \frac{q}{W} \cdot F(W^2)$, where $W$ is the center-of-mass 
(c.m.) total energy and $q$ is the c.m. momentum of the final-state particles. The factor $F(W^2)$, 
which does not vanish at threshold, comes from the sum of production amplitudes squared, and 
$\frac{q}{W}$ from the integration over the final-state phase space. Because $W^2$ is linearly related 
to the photon energy $E_\gamma$ for the charmonium photoproduction, the value 
of $\sigma_t$
%$\sigma_t^2$ 
as a function 
of the photon energy in the laboratory frame $E_\gamma$ reaches zero at the threshold energy $E_\gamma = 
E_\gamma^{th}$ without any singularity (i.e., if the final-state $S$-wave does not vanish at threshold).

%The results of this work for $\sigma_t^2(\gamma p\to J/\psi p)$
%are shown as a function of $E_\gamma$ in Fig.~\ref{fig:fig1}(a). 
%The fit of the $J/\psi$ $\sigma_t^2(E_\gamma)$ data from the 
%GlueX Collaboration with the formula:
%\begin{equation}
%	\sigma_t^2(E_\gamma) = b_1\delta + b_2\delta^2 
%	+ b_3\delta^3 + b_4\delta^4
%        \label{eq:eq1}
%\end{equation}
%with one of four free parameters included in $\delta = E_\gamma 
%- E_\gamma^{th}$, is shown in Fig.~\ref{fig:fig1}(a) by a solid 
%red line. For the parameter $E_\gamma^{th}$, the fit results in 
%the value ($8.21\pm 0.18$)~GeV, corresponding to the mass 
%$m_{J/\psi} = (3.097\pm 0.010)~GeV/c^2$. It is in good agreement 
%with the RPP value $m_{J/\psi} = (3096.900\pm 
%0.006)~MeV/c^2$~\cite{Tanabashi:2018oca}.  Although the estimate 
%made here for the charmonium mass cannot compete in precision 
%with the known RPP value, the agreement observed indicates the 
%good quality of the GlueX data and the correctness of the 
%photon-beam energy calibration, the systematic uncertainty in 
%which was determined as $2\times 10^{-3}$~\cite{Somov:2019}.

\begin{figure}[htb]
\begin{center}
\includegraphics[height=4in, keepaspectratio, angle=90]{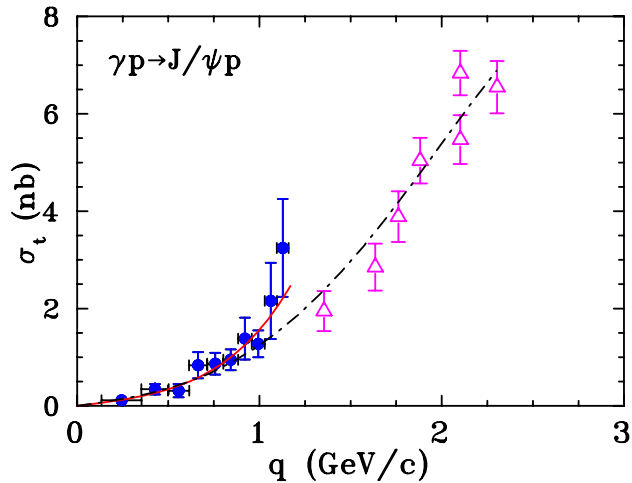}
\end{center}

\vspace{-5mm}
\caption{Results of the GlueX work (blue solid circles) for the $\gamma p\to J/\psi p$ total cross sections
  $\sigma_t$~\protect\cite{Ali:2019lzf} are shown as a function of the c.m. momentum $q$ of the final-state
  particles. The previous SLAC data~\protect\cite{Camerini:1975cy} (where the determination of $\sigma_t$
  from the differential cross sections was performed in Ref.~\protect\cite{Ali:2019lzf}) are shown by the
  magenta open triangles. The vertical error bars represent the total uncertainties of the results (summing
  statistical and systematic uncertainties in quadrature). The horizontal error bars for the GlueX data 
  reflect the energy binning. The red solid curve shows the fit of the GlueX data with 
  Eq.~(\protect\ref{eq:eq2}), while the black dot-dashed curve shows a combined fit of the GlueX and SLAC 
  data with Eq.~(\protect\ref{eq:eq2}).} \label{fig:fig1}
\end{figure}

Traditionally, the $\sigma_t$ behavior of a binary inelastic reaction near threshold can be described 
as a series of odd powers of $q$. In the energy range under study,
\begin{equation}
        \sigma_t(q) = a_1q + a_3q^3 + a_5q^5
        \label{eq:eq2}
\end{equation}
is enough to describe the near threshold cross section $\sigma_t(q)$ quite well. The fit of the 
GlueX data with Eq.~(\ref{eq:eq2}) is shown in Fig.~\ref{fig:fig1} by a solid red curve. The fit 
of both GlueX and SLAC data is shown in Fig.~\ref{fig:fig1}, as well, by a black dot-dashed 
curve. The best-fit results are summarized in Table~\ref{tbl:tab1}.  Note that the SLAC experiment 
measured the differential cross-section $d\sigma/dt$ at $t=t_{min}$ as a function of $E_{\gamma}$. 
To calculate the total cross sections from the SLAC data, we have used a dipole $t$-dependence as 
was done in Ref.~\cite{Ali:2019lzf}.
%------------------------------------------
\begin{table}[htb!]

\centering \protect\caption{The fit of the 
	GlueX~\protect\cite{Ali:2019lzf} (2nd colomn) and GlueX with 
	SLAC~\protect\cite{Camerini:1975cy} (3rd colomn) data with 
	Eq.~(\protect\ref{eq:eq2}). Error bars of the GlueX data 
	represent the total uncertainties (summing statistical and 
	systematic uncertainties in quadrature).} 
\vspace{2mm}
{%
\begin{tabular}{|c|c|c|}
\hline
$a_i$                  & GlueX Data     &  GlueX and SLAC Data \\
\hline
$a_1$ [nb/(GeV/c)]     &  0.46$\pm$0.16 &  0.53$\pm$0.12 \\
$a_3$ [nb/(GeV/c)$^3$] &  0.83$\pm$0.91 &  0.78$\pm$0.16 \\
$a_5$ [nb/(GeV/c)$^5$] &  0.28$\pm$0.87 & -0.06$\pm$0.03 \\
\hline
$\chi^2$/dof           &  0.67          & 0.98           \\
\hline
\end{tabular}} \label{tbl:tab1}
\end{table}
%------------------------------------------

The parameter $a_1$ as obtained from the fit of the GlueX data alone and both the GlueX 
and SLAC data agree within the uncertainties (Table~\ref{tbl:tab1}), the near threshold 
behavior is very similar for both curves. The linear term in Eq.~(\protect\ref{eq:eq2}) 
is determined here by the $S$-waves only (with total spin $1/2$ and/or $3/2$). The 
contributions to the cubic term come from both the $P$-wave amplitudes and the $W$ 
dependence of the $S$-wave amplitudes, and the fifth-order term arises from the $D$-waves 
and the $W$ dependencies of the $S$- and the $P$-waves.
%-------------------------------------------------
\vspace{10mm}
%-------------------------------------------------

The $\sigma_t(\gamma p\to J/\psi p)$ data is related to the $J/\psi N$
scattering length, $\alpha_{J/\psi p}$, by the threshold relation: 
\begin{equation}
	\frac{d\sigma(J/\psi p\to J/\psi p)}{d\Omega} 
	|_{th} = |\alpha_{J/\psi p}|^2.
        \label{eq:eq7}
\end{equation}
In fact, the cross-section includes contributions from two independent $S$-wave scattering lengths 
with total spins $1/2$ and $3/2$. 

In the VMD framework, $\alpha_{J/\psi p}$ appears also in the $\gamma p\to J/\psi p$ cross-section near 
threshold~\cite{Titov:2007xb}:
\begin{equation}
	\sigma_t(\gamma p\to J/\psi p)|_{th} = 
	\frac{q}{k} \cdot \frac{4\alpha\pi^2}{g_{J/\psi \gamma}^2}
	\cdot |\alpha_{J/\psi p}|^2,
        \label{eq:eq3}
\end{equation}
where $k$ is the c.m. momentum of the incident photon at the $\gamma p\to J/\psi p$ threshold, $\alpha$ is
the fine-structure constant, $M$ is the $J/\psi$ mass, and $g_{J/\psi \gamma} = 5.58\pm 0.07$ is the
$\gamma \to J/\psi$ coupling, as determined from the $J/\psi\to e^+e^-$ decay width~\cite{Anashin:2018ldj}. 
This result came recently from the KEDR Collaboration that determined $\Gamma(J/\psi\to e^+e^-)$ using
the KEDR detector at the VEPP-4M $e^+e^-$ collider. Summing the statistical and systematic uncertainties in
quadrature, one gets $\Gamma(J/\psi\to e^+e^-) = (5.55\pm 0.11)$~keV.  
%That can be converted in to 
%dimensionless coupling $g_{J/\psi \gamma} = \frac{M(J/\psi)^2}
%{g_{J/\psi}}$.
As known in VMD, the coupling $g_{J/\psi \gamma}$ is related to the electron width of the vector meson 
% $J/\psi$ in our case, 
$\Gamma(J/\psi\to e^+e^-)$ as:
\begin{equation}
        g_{J/\psi \gamma} = \sqrt{\frac{\pi\alpha^2M}
        {3\Gamma(J/\psi\to e^+e^-)}}.
% = 11.15\pm 0.10.
%\begin{equation}
%        g_{J/\psi \gamma} = \sqrt{\frac{3\times M(J/\psi)^3\times
%        \Gamma(J/\psi\to e^+e^-)}{(4\pi\alpha^2)}},
        \label{eq:eq4}
\end{equation}
%where $\Gamma(J/\psi\to e^+e^-)$ = 
%$\Gamma(J/\psi)_{tot} \times BR(J/\psi\to e^+e^-)$. %\textcolor{red}
%
%Then, $g_{J/\psi \gamma}$ coupling is a dimensionless parameter 
%is also introduced according to relation: 
%$g_{J/\psi \gamma} = \frac{M(J/\psi)^2}{g_{J/\psi}}$. 
%
%One can evaluate the $g_{J/\psi \gamma}$, according to
%Eq.~(\ref{eq:eq4}):
%\begin{equation}
%        g_{J/\psi \gamma} = (0.860 \pm 0.008)~GeV^2,
%        \label{eq:eq5}
%\end{equation}
%\textcolor{red}{Denis, please check this results.}
%here, the dominant uncertainty comes from the uncertainty of
%$\Gamma(J/\psi\to e^+e^-)$.
%
%The factor $2/(3+q^2/M^2)$ in Eq.~(\ref{eq:eq3}) takes into account 
%the ratio of the polarization states for the $\gamma$ and $J/\psi$ 
%initial states~\cite{Kubarovsky:2015aaa} in case of s-wave, which 
%is $2/3$ at threshold.

Combining Eq.~(\ref{eq:eq3}) with the $a_1$ value from fitting Eq.~(\ref{eq:eq2}) 
to the GlueX $\sigma_t(\gamma p\to J/\psi p)$ data given in Table~\ref{tbl:tab1} 
results in
\begin{equation}
	|\alpha_{J/\psi p}| = \frac{g_{J/\psi \gamma}}{2\pi}
	\sqrt{\frac{ka_1}{\alpha}} = (3.08\pm 0.55)~{\rm mfm} ,
        \label{eq:eq6}
\end{equation}
which should be considered just as an estimate assuming only the sequence $\gamma\to J/\psi$,
$J/\psi p\to J/\psi p$. 
%A more detailed analysis is needed, however, to exclude contributions from $\gamma\to\rho^0$ 
%and $\rho^0p\to\rho^0p$ that contain, in particular, $\pi^0$ exchange, and from the similar 
%transitions $\gamma\to\omega$ and $\gamma\to\phi$. 
Taking into account the overall systematics of the GlueX data of 27\%, we obtain finally
$|\alpha_{J/\psi p}| = (3.08\pm 0.55 ({\rm stat.}) \pm 0.42 ({\rm syst.}))$~mfm.

To estimate the theoretical uncertainty related to the VMD model, we refer back to 1977, when 
Boreskov and Ioffe~\cite{Boreskov:1976dj} estimated the cross section of $J/\psi$ 
photoproduction in a peripheral model and found a strong energy dependence close to threshold 
because the non-diagonal $\gamma p\to J/\psi p$  must have larger transfer momenta 
versus elastic scattering. This results in a violation of VMD by a factor of 5 or so. In 1993, 
Boreskov and co-workers showed that a fluctuation of a photon into open charm is preferable 
than into a $J/\psi$~\cite{Boreskov:1992ur}.  In addition, in Eq.~(\ref{eq:eq3}) we have not 
included a factor introduced in the VMD model in Ref.~\cite{Kubarovsky:2015aaa}, which takes 
into account the difference between polarization degrees of freedom in the $\gamma p\to J/\psi 
p$ and $pJ/\psi\to pJ/\psi$ reactions. Such a factor that equals $2/3$ at threshold for the 
$S$-wave has not been used in the previous analysis of the scattering length and we consider 
it as a systematic uncertainty related to the VMD model.

\begin{figure}[htb]
\begin{center}
\includegraphics[height=4in, keepaspectratio, angle=90]{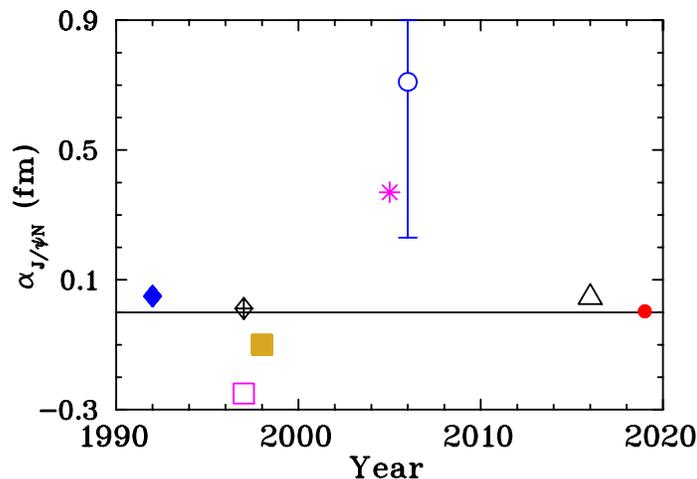}
\end{center}

\vspace{-5mm}
\caption{Comparison of different determinations of $\alpha_{J/\psi p}$ scattering length. The red 
  filled circle shows our result,
  the black open triangle is from Ref.~\protect\cite{Gryniuk:2016mpk}, the blue open circle is 
  from Ref.~\protect\cite{Yokokawa:2006td},
  the magenta star is from Ref.~\protect\cite{Sibirtsev:2005ex},
  the yellow filled square is from Ref.~\protect\cite{Hayashigaki:1998ey},
  the magenta open square is from Ref.~\protect\cite{Brodsky:1997gh},
  the black open diamond with cross is from Ref.~\protect\cite{Shevchenko:1996ch}, and
  the blue filled diamond is from Ref.~\protect\cite{Kaidalov:1992hd}.}
  \label{fig:fig3}
\end{figure}

Nevertheless, the present estimate for $|\alpha_{J/\psi p}|$ using the near-threshold 
photoproduction of charmonium data from the GlueX Collaboration is within the broad range 
defined by other $\alpha_{J/\psi p}$ values available in the literature: 
%------------
0.046$\pm$0.005~fm from a global fit to both previous differential and total cross section 
data~\cite{Gryniuk:2016mpk}, 
%------------
0.37~fm from the multipole expansion and low-energy theorems in QCD~\cite{Sibirtsev:2005ex},
%------------
-0.25~fm from the gluonic van der Waals interaction~\cite{Brodsky:1997gh},
%------------
0.05~fm from a multipole expansion for heavy-quarkonia interactions with gluon fields and 
low-energy QCD theorems for gluon interactions with nucleons~\cite{Kaidalov:1992hd},
%------------
0.71$\pm$0.48~fm from Lattice QCD calculations~\cite{Yokokawa:2006td}, 
%------------
-0.1~fm from QCD sum rules~\cite{Hayashigaki:1998ey}, and
%------------
0.012~fm from the gauge-invariant quark-antiquark Greens function for the expression of 
the non-relativistic meson scattering amplitude on the external gluon 
field~\cite{Shevchenko:1996ch} (see Fig.~\ref{fig:fig3}).
%-------------------------------------------------
%-------------------------------------------------

Our result disagrees with the recent work of Ref.~\cite{Gryniuk:2016mpk} 
based on the fit of the previous data from SLAC and other data far away
from the threshold~\cite{Camerini:1975cy}. Note that our value of 
$|\alpha_{J/\psi p}|$ is much less than the recent $\omega$ photoproduction at the threshold result from 
the A2 Collaboration at MAMI for the $\omega p$ scattering length $|\alpha_{\omega p}|$ = 
(0.82$\pm$0.03)~fm~\cite{Strakovsky:2014wja} and much less than a typical size of a hadron.

%--------------- Summary ----------------------------------
%\section{Summary}
%\label{sec:sum}

In summary, an experimental study of charmonium photoproduction off the proton was conducted by the 
GlueX Collaboration at JLab ~\cite{Ali:2019lzf}. The proximity of the new GlueX data to the threshold 
allows to estimate the $J/\psi p$ scattering length within the VMD model.  Note that the first data bin 
(Fig.~\ref{fig:fig1}) has a weighted average of $q = 230$~MeV/c, or $1/q = 0.86$~fm is of the order of 
the proton size.  Our result for the $|\alpha_{J/\psi N}|$ scattering length disagrees with previous 
theoretical results individually, though it is within the wide range of these predictions.  Our small 
value of scattering length vs a typical size of a hadron,~1~fm, indicates that the proton is 
transparent for $J/\psi$. Within VDM the $J/\psi$ photo production is suppressed as $m_\omega^2/m_{J/\psi}^2$
as compared with the $\omega$ photoproduction.
%The $J/\psi p$ interaction decreases as $(m_\omega / M)^2$ = 1/20 to compare with $\omega$ photoproduction 
%or just compare $\omega$ photoproduction and $\omega - J/\psi$ mixing.
Present and future experiments at Jefferson Lab that are aimed to measure the charmonium production  off 
proton and nuclei~\cite{jlab1,jlab2,jlab3,jlab4} will allow further studies of the $J/\psi N$ interaction 
and will give also access to a variety of other interesting physics aspects that are present in the 
near-threshold region. 

%-------------- Acknowledgments ----------------------------
%\section{Acknowledgments}

We thank Konstantin~Boreskov and Arkady Vainstein for useful remarks and continuous interest in the paper 
and Daniel~Carman, Eugene Chudakov, and Michael Eides for valuable comments. This work was supported in 
part by the U.~S.~Department of Energy, Office of Science, Office of Nuclear Physics under Award No. 
DE--SC0016583 and Contract No. DE--AC05--06OR23177. One of the authors (I.S.) highly appreciates the 
hospitality extended to him by the Munich Institute for Astro- and Particle Physics (MIAPP) of the DFG 
cluster of excellence ``Origin and Structure of the Universe".

%---------------------- REFERENCES ----------------------

%-----------------------------------------------
\end{document}